# Quantum Algorithms for Network Signal Coordination

Prof. Vinayak Dixit and Richard Pech

**Abstract**
There has been increasing interest in developing efficient quantum algorithms for hard classical problems. The Network Signal Coordination (NSC) problem is one such problem known to be NP complete. We implement Grover's search algorithm to solve the NSC problem to provide quadratic speed-up. We further extend the algorithm to a Robust NSC formulation and analyse its complexity under both constant and polynomial-precision robustness parameters. The Robust NSC problem determines whether there exists a fraction ($\alpha$) of solutions space that will lead to system delays less than a maximum threshold ($K$). The key contributions of this work are (1) development of a quantum algorithm for the NSC problem, and (2) a quantum algorithm for the Robust NSC problem whose iteration count is $O(1/\sqrt{\alpha})$, independent of the search space size, and (3) an extension to polynomial-precision robustness where $\alpha = \alpha_0/p(N)$ decays polynomially with network size, retaining a quadratic quantum speedup. We demonstrate its implementation through simulation and on an actual quantum computer.

## 1. Introduction

Research in quantum computing and algorithms over the past three decades have theoretically demonstrated the potential gains through "quantum speedup" (Montanaro, 2016). At a fundamental level, quantum computers differ from classical computers in their ability to leverage quantum mechanical properties such as superposition, entanglement and interference to speedup computations.

There has been groundbreaking theoretical work demonstrating that quantum algorithms relying on quantum logic gates can provide significant speedups to classical computers (Deutsch and Jozsa, 1992; Simon, 1994; Shor, 1994; Dürr and Høyer, 1996; Bernstein and Vazirani, 1997; Grover, 1997) One of the most celebrated is the Shor's (1994) algorithm, that demonstrated that quantum computers can solve the prime factorization problem exponentially faster than classical computers, having significant implications on cryptography. Though subject to some debate, recently "Quantum Supremacy" was demonstrated on a problem that would take a classical supercomputer 10,000 years to be completed by 53 qubit Sycamore processor in 200 seconds (Arute et al. 2019). Even though quantum computers might outperform classical computers by orders of magnitude, Aaronson (2008) identified NP-Complete problems as one of the limits of Quantum Computers.

Urban traffic signal control is a complex and critical problem in transportation engineering. Optimizing the timing of traffic lights across a network of intersections can significantly reduce congestion, travel delays, and emissions. However, this problem is known to be computationally hard. Gartner et al. (1975) introduced a mixed-integer approach for the "Network Coordination Problem". Dauscha et al. (1985) showed an analogous Cyclic Schedule Problem to be NP-complete. Serafini and Ukovich (1989), Hassin (1996) as well as Wunsch (2008) proved the NSC problem to be NP-complete. The proposed problem was shown by Hassin (1996) to be NP-complete by reducing it to a minimum cluster problem, which is equivalent to the max cut problem (Garey and Johnson 1978).

## 2. Network Signal Coordination

The Network Signal Coordination problem is represented as a graph problem. Consider a Graph $G = (V, A)$, the mathematical program for this can be represented as:

$$\min \sum_{(i,j) \in A} h_{ij}(\mu_i - \mu_j), \quad [1]$$

where the delay function $h_{ij}(\cdot)$ is periodic with a period $C$, such that $h_{ij}(\mu_i - \mu_j) = h_{ij}(\mu_i - \mu_j \pm C)$ for all offsets $\mu_i, \mu_j$ $\forall i, j \in N$ and $(i,j) \in A$. In complexity theory, optimization problems are reduced to an equivalent decision problem. In the case of the NSC optimization problem the decision problem involves determining whether there exists a set of offsets $\mu$, that satisfies the following condition:

$$\exists \mu \in N^V \mid \sum_{(i,j) \in A} h_{ij}(\mu_i - \mu_j) \leq K, \quad K \in N^+ \quad [2]$$

The optimization problem presented in Equation 1 and the Decision Problem in Equation 2 have been framed with node offset variables. In this paper, we develop a Quantum Oracle for this decision problem.

However, there are equivalent formulations relying on arc offset variables (see for example, Gartner and Little 1975, Gartner et al. 1975, Importa and Sforza, 1982 and Wunsch, 2008). An additional constraint needs to be included to ensure that the sum of offsets over any cycle within the graph should be an integer multiple of the Signal Cycle (Period) Length $C$. Given that every cycle within a Graph can be generated using a subset of the Cycle Basis ($\Xi$) of a graph, which has a size of $|A| - |V| + 1$, cycles. We need only $|A| - |V| + 1$ constraints to satisfy the cycle constraint. Paton (1969) presented an efficient classical algorithm that was polynomial in time and space needed as a function of the number of nodes ($|V|$) in the graph to determine the full set of the cycle basis ($\Xi$), that has $|A| - |V| + 1$ elements. Therefore, given a set of offsets $\phi$, the feasibility can be checked in $O(|A| \cdot (|A| - |V| + 1))$. However, finding the optimal solution was still an NP complete problem, requiring checking the entire solution space.

To develop an efficient quantum algorithm, we need to create Quantum Oracle for the total delay $(\sum_{(i,j) \in A} h_{ij}(\mu_i - \mu_j))$ for the Network Signal Coordination Problem. The node-based formulation shown in Equations 1-2 are more appealing to develop an oracle for as they have no additional periodicity constraint as shown in Equation 4. Without loss of generality, we use the node-based formulation for the rest of the manuscript.

### 2.1 Feasibility Function

The definition of the Network Coordination Problem requires that the decision condition in Equation 2 be satisfied. Therefore, for a given set of offsets $\mu \in N^V$ we construct a feasibility function $\kappa$ for Equation 2, such that,

$$\kappa(\mu) = \begin{cases} 1 & if \sum_{a \in A} h_a(\mu_i - \mu_j) \leq K \quad \forall \mu_i \in \mu \text{ and } (i,j) \in A \\ 0 & else \end{cases} \quad [3]$$

This will be implemented in a quantum routine to solve the NSC problem.

## 2.2 Quantum Algorithm for Network Signal Coordination

The quantum routine for the Network Coordination Problem relies on the Grover's algorithm (Grover, 1997 and see Nielsen and Chuang, 2019) to search if there are solutions that provide total delays that are less than a given threshold. The overall approach involves implementing the Oracle and then implementing a Grover's search via the feasibility function. The algorithm is outlined below in Figure 1.

1) **Initialize**: The offsets of each node are represented by a quantum register, which are initialized into a uniform superposition, by applying a Hadamard gate to each of the node registers. We have three additional registers, the edge delay ($d$), the register storing the sum ($s$) and a feasibility register ($f$).

$$\frac{1}{\sqrt{C^{|V|}}} \sum_{x=0}^{C^{|V|}-1} |\mu_0, \mu_1 \ldots \mu_{|V|-1}\rangle \, |0\rangle|0\rangle|0\rangle \qquad [4]$$

2) Apply delay oracles for each of the edges $(i,j) \in A$, given by $h_{ij}(\mu_i - \mu_j)$. This is done individually, and all the delay outputs are stored in their individual registers.

$$\frac{1}{\sqrt{C^{|V|}}} \sum_{x=0}^{C^{|V|}-1} |\mu_0, \mu_1 \ldots \mu_{|V|-1}\rangle |h_{|A|-1}\rangle|0\rangle|0\rangle \qquad [5]$$

3) **Total Delay**: This routine adds the delays associated with edge and is done sequentially.
   a. Apply arithmetic adder gadgets to sum the associated delay edge registers, which is to be stored in the sum register. As quantum adder gates are used, this is made simple by appropriating the operation in terms of binary operations. Each of the delay output results are temporarily stored into a form that allows for an $n$-bit addition between the temporarily formatted delay output and the sum register. That is, the delay output gets ancillary 0 qubits appended in front so that the total number of qubits matches with the sum register. Due to the nature of addition, a carry qubit should be included for completeness. Adding to this, the size of the sum register can be found beforehand by noting the number of edges of the graph problem. For the first edge the summation with 0 leads to the delay from the first edge

$$\frac{1}{\sqrt{C^{|V|}}} \sum_{x=0}^{C^{|V|}-1} |\mu_0, \mu_1 \ldots \mu_{|V|-1}\rangle |h_{|A|-1}\rangle|h_0\rangle|0\rangle \qquad [6]$$

   b. To add delays from all edges, repeat, Steps 2.1 and 2.2 for all edges ($i \in A$).

$$\frac{1}{\sqrt{C^{|V|}}} \sum_{x=0}^{C^{|V|}-1} |x_0, x_1 \ldots x_{|A|-1}\rangle|h_{|A|-1}\rangle \left|\sum_{i \in A} h_i\right\rangle |0\rangle \qquad [7]$$

   In the case of one qubit, the individual adder operations can be done by applying a CNOT gate on the delay output, then applying an adder on the formatted delay output and the sum and then inverting the CNOT gate so that only the sum register changes. The size of the sum register (and the number of ancillary qubits to format the delay outputs) would be $\varepsilon = \text{floor}(log_2(|A|)) + 1$. An example circuit to demonstrate this is given below. Half adders are used in the code used to execute the algorithm, as it avoids extra qubits needed for the addition.

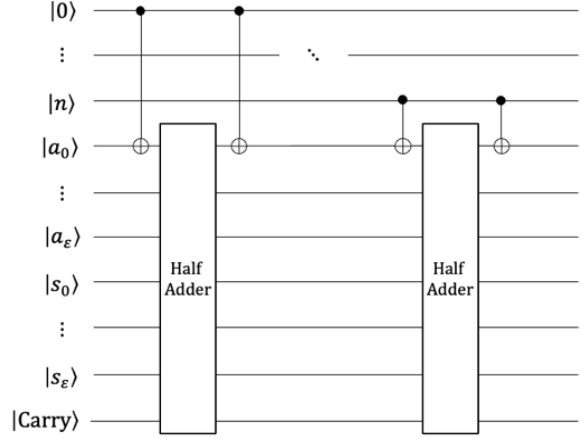

**Figure 1: Quantum Circuit for Sequential Addition**

4) **Feasibility Function**: To determine the feasible offsets such that the sum of delays being less than the threshold $K$, we apply an integer comparator gate that compares the sum operator to some given integer value. This flips the register to 1 if the threshold condition is satisfied.

$$\frac{1}{\sqrt{C^{|V|}}} \sum_{x=0}^{C^{|V|}-1} |x_0, x_1 \ldots x_{|A|-1}\rangle |h_{|A|-1}\rangle \left|\sum_{a \in A} h_a\right\rangle |\kappa\rangle \qquad [8]$$

5) **Grover's Search:** We then apply Grover's algorithm to find input configurations that satisfy the threshold condition. This requires a phase flip on the feasibility register to 'mark' the good states and then applying the inverse of all the previous operations to return everything back to 0, except for the node registers. A Grover diffuser is then applied to amplify the marked good states, which acts as an efficient method to search for solutions. It is important to note that within each Grover iteration, the oracle must be applied and then fully uncomputed (reversed) before the diffuser is applied. Specifically, after the phase flip on the feasibility register, all intermediate computations—the delay oracle evaluations, the sequential addition, and the comparator—must be run in reverse to restore the ancilla registers to their initial states. This uncomputation step is necessary because Grover's algorithm requires the oracle to act as a phase oracle on the node registers alone, with no entanglement remaining with ancillary qubits. Consequently, the effective circuit depth per Grover iteration is approximately twice the oracle depth, and all intermediate registers must be preserved (not measured) during the computation.

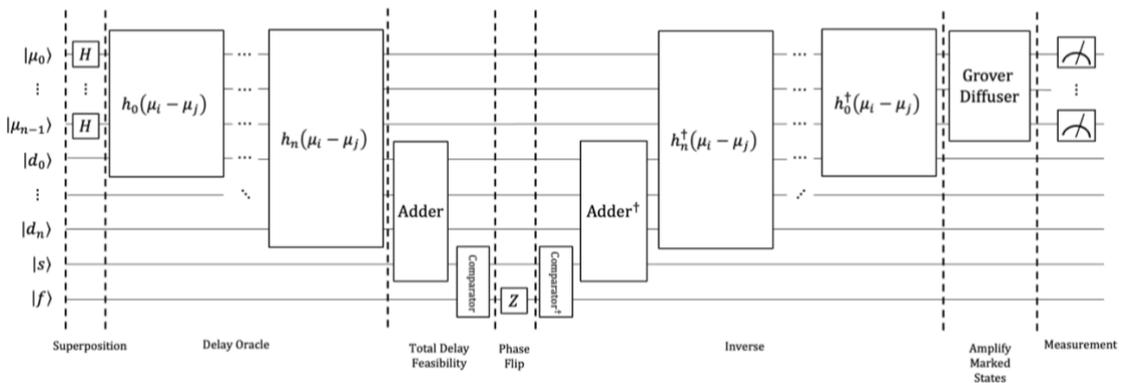

**Figure 2: Quantum Circuit for the Robust Network Signal Coordination Problem**

## 2.3 Complexity

**Proposition 1 (Quadratic Speedup for NSC).** For the NSC decision problem on a graph G = (V, A) with cycle length C, the Grover-based quantum algorithm finds a feasible offset assignment (if one exists) using $O\left(C^{\frac{|V|}{2}}\right)$ oracle queries, compared to $O(C^{|V|})$ for classical exhaustive search. Each oracle query has gate complexity O(|A| log |A|).

The sequential Adder in Step 2 has a complexity of $O(|A|^2)$, as the Adder needs to be carried out for each edge. Specifically, each of the |A| additions operate on a sum register of size ε = ⌊log₂|A|⌋ + 1, with each addition requiring O(ε) gates using a half-adder chain, giving a total complexity of O(|A| × ε) = O(|A|²). If the offset on each vertex of the graph is represented by $n$ qubits, then the node registers alone require $n|V|$ qubits. In addition to these node register qubits, the full circuit requires |A| qubits for the delay registers (one per edge), $\lfloor log_2|A|\rfloor + 1$ qubits each for the sum register and ancilla register, one carry qubit, one feasibility register qubit, and $\lfloor log_2|A|\rfloor$ comparator ancilla qubits. The total qubit count is therefore approximately n|V| + |A| + 3⌊log₂|A|⌋ + 4. However, the most expensive part of the algorithm is Grover's search, which provides a quadratic speedup $2^{n|V|/2}$, where $2^{n|V|}$ is the size of the total search space. Note that this is the number of Grover iterations; each iteration requires executing the full oracle circuit, so the total gate complexity is $O(\sqrt{N} \times C_{oracle})$, where $C_{oracle}$ is the cost of a single oracle evaluation.

## 3. Background and Robust NSC Formulation

The NSC problem becomes even more challenging due to the dynamic and stochastic nature of traffic flow and various uncertainties, such as fluctuations in demand, incidents and weather. A robust traffic signal control problem seeks to optimize the performance of a traffic signal system (e.g., minimize delay, emissions, or queue length) while ensuring reliability and effectiveness across a range of uncertain conditions. Traditional signal systems often rely on simplified models or fixed timings that cannot adapt well to real-world variability (Wei et al., 2021). Recent advancements include use of robust optimization techniques that account for uncertainty. The Robust NSC optimization problem has been formulated as:

1. **The minimax formulation** is represented as $\min_{x \in X} \max_{u \in U} f(x, u)$, where, $x$ is the signal control variables (e.g., green times, phase orders), $u$: Uncertain traffic parameters (e.g., arrival rates, turning ratios), $f(x, u)$ is the Objective function (e.g., total network delay). This formulation finds the signal plan $x$ that minimizes the worst-case objective value over the uncertainty set $U$. (Farokhi & Langbort, 2013; Osorio & Chong 2015)

2. **Two-Stage Robust Optimization** is a commonly used approach that involves, a Stage 1, that determines an initial plan based on known historical or nominal data, and a Stage 2 that adjusts or refines the plan in response to real-time uncertainty. These are broadly formulated as a bi-level mixed-integer linear program (MILP) (Han et al. 2015; Zheng et al. 2023; Qi et al. 2023), to find robustly optimal solutions to the variability.

3. **Chance Constrained optimization** is represented as $\min_{x \in X} f(x)$, $\Pr\{h(x) \leq T\} \geq \eta$, is useful for robust optimality under stochastic or noisy conditions. optimizes traffic signals to minimize objectives like delay while ensuring that critical constraints (e.g., maximum queue lengths) are satisfied with high probability (e.g., 95%), allowing for a small, controlled risk of constraint violation under uncertain traffic conditions. Unlike strict robust optimization, which prepares for worst-case scenarios that tend to be overly conservative, chance constraints balance performance and reliability, making them more practical for urban traffic management. Techniques usually apply sample-based methods to approximate the constraint satisfaction. Kouvelas et al. (2014), applied chance-constrained model predictive control (MPC) for real-time signal control, Zhao and Zhang (2018), formulated it for adaptive traffic signal control, and Kovács et al. (2016), extended their 2014 work into distributed MPC for large urban networks.

4. **Metaheuristics** such as Genetic Algorithms (GA), Particle Swarm Optimization (PSO), and Ant Colony Optimization (ACO) are widely used due to their flexibility. A Vancouver-based case study applied GA to three congested intersections and achieved a 25% reduction in vehicle delays and a 12% reduction in emissions (Deshpande, 2024). Jovanović and Teodorović (2017) applied Bee Colony Optimization to isolated intersections with effective results.
5. **Reinforcement Learning (RL)**, particularly Deep RL, has gained prominence for its ability to learn from interaction with traffic environments. Wei et al. (2021) survey recent RL models and evaluation methods, demonstrating the scalability and adaptability of RL-based systems. Graph neural networks have enabled coordination among RL agents across networks. The CoLight model uses a graph attention network to optimize signal coordination in multi-agent RL settings (Ge et al., 2022). Robust RL techniques that expose agents to perturbations during training improve resilience to real-world variability (ORoojlooy et al., 2020). Hybrid methods combine the strengths of multiple approaches. Dixit et al. (2020) and Yang et al. (2024) integrated Max-Pressure control with Deep RL to dynamically manage signal phases and durations.

In the quantum computing domain, variational quantum algorithms such as the Quantum Approximate Optimization Algorithm (QAOA) and the Variational Quantum Eigensolver (VQE) have been applied to combinatorial optimization problems including MaxCut and vehicle routing. Quantum annealing approaches (e.g., on D-Wave hardware) have also been explored for traffic signal optimization. However, these approaches target different problem formulations and do not provide the provable worst-case speedup guarantees offered by Grover-based methods. The present work is, to our knowledge, the first to apply Grover's algorithm specifically to the NP-complete Network Signal Coordination problem with formal complexity analysis.

The robust NSC is formulated as a chance constrained optimization problem, where in, the optimal values are selected such that there is sufficient redundancy in the design domain. That is:

$$\max_{S \subset C^{|V|}} |S| \quad | \quad \sum_{(i,j) \in A} h_{ij}(\mu_i - \mu_j) \leq K \quad \forall \mu \in S \qquad [9]$$

Where $\mu = \{\mu_i\}_{\forall i \in V}$ is the solution vector. The equivalent decision problem is:

$$\exists\, S \quad | \quad S = \left\{ \mu : \sum_{(i,j) \in A} h_{ij}(\mu_i - \mu_j) \leq K,\ \mu \in C^{|V|} \right\} \text{ and } |S| \geq \delta \qquad [10]$$

Essentially Equation 10 states that the number of elements in the set of feasible offsets $S$ that have a total delay less than $K$ should be greater than $\delta$. For the special case, where $\delta = 1$, the problem reduces to the standard NSC problem which is known to be NP-complete and solved using methods described in Section 2. However, for the case where $\delta = \alpha 2^n$, where $0 < \alpha < 1$ provides a robustness index on the percentage of the feasible set that should be under the threshold.

### 3.1 Complexity

**Proposition 2 (Constant-α Robust NSC).** For the Robust NSC decision problem with $\delta = \alpha C^{|V|}$ where $\alpha \in (0,1)$ is a fixed design parameter, the quantum algorithm requires $O(1/\sqrt{\alpha})$ Grover iterations to find a feasible solution. Each iteration has cost $C_{oracle}$, giving total query complexity $O(C_{oracle}/\sqrt{\alpha})$, which is independent of the search space size.

To find the solution with a high probability the oracle and Grover diffuser (steps 2 to 5) needs to be run to $\lceil \pi\sqrt{N}/4 \rceil$ times (Grover, 1997). However, if we need to find whether there are at least $\delta$ solutions, we will need to run the Grover's operator $\lceil \pi\sqrt{N/\delta}/4 \rceil$, which is equivalent to $\lceil \pi\sqrt{1/\alpha}/4 \rceil$. As $\alpha$ is a constant design parameter of the robust optimization problem, the number of Grover iterations is $O(1/\sqrt{\alpha})$. That is, the number of Grover iterations required is independent of the search space size N, depending only on the design parameter α. This is a notable result, as it means the iteration count does not grow with the network size. However, each iteration still requires executing the full oracle circuit, whose gate complexity grows polynomially with the graph size. Section 3.2 extends this analysis to the more realistic setting where α is not constant but decays polynomially with the network size.

It is important to note that the Robust NSC decision problem (Equation 10) asks whether $|S| \geq \delta$, which is a counting question rather than a search question. Grover's algorithm finds solutions but does not directly count them: finding one feasible offset proves $|S| \geq 1$ but not $|S| \geq \delta$. To properly answer the decision problem, one should employ quantum counting (Brassard et al., 2002), which combines the Grover operator with quantum phase estimation to estimate the number of marked states M. Quantum counting estimates M to within additive accuracy ε using $O(\sqrt{N}/\varepsilon)$ queries, which for $\varepsilon = \delta/2 = \alpha N/2$ gives $O(\sqrt{(2/\alpha)}) = O(1/\sqrt{\alpha})$ queries — a bound that, like the search complexity, is independent of the search space size N when α is constant.

Furthermore, the quantum advantage for the Robust NSC should be compared against the classical baseline. If a fraction α of all offset configurations are feasible, then classical random sampling succeeds with probability α per trial, requiring $O(1/\alpha)$ random samples to find a feasible solution with high probability. The quantum algorithm improves this to $O(1/\sqrt{\alpha})$ Grover iterations — a quadratic speedup in the sampling step. However, when α is treated as a constant (fixed design parameter), both the classical $O(1/\alpha)$ and quantum $O(1/\sqrt{\alpha})$ algorithms require a constant number of samples with respect to the search space size N. The quantum advantage is therefore the quadratic improvement from $O(1/\alpha)$ to $O(1/\sqrt{\alpha})$ in the number of oracle calls, which is meaningful when each oracle evaluation is expensive or when α is small.

### 3.2 Polynomial-Precision Robustness

The analysis above treats α as a fixed design parameter, independent of the network size. However, in realistic traffic networks, as the number of intersections |V| and links |A| grows, the fraction of offset configurations satisfying the delay threshold K typically decreases—more constraints must be satisfied simultaneously, and the feasible region of the solution space shrinks. This motivates a more general analysis where α depends on the problem size.

Let $N = C^{|V|}$ denote the search space size. We model the polynomial decay of feasibility as $\alpha(N) = \alpha_0/p(N)$, where $\alpha_0 \in (0,1)$ is a base robustness parameter and $p(N)$ is a polynomial function of N (or equivalently of |V|, |A|, or C). The number of feasible solutions is then $\delta = \alpha N = \alpha_0 N/p(N)$.

Theorem 1 (Quantum Speedup for Polynomial-Precision Robustness). For the Robust NSC problem with search space size $N = C^{\wedge}|V|$ and feasibility fraction $\alpha(N) = \alpha_0/p(N)$ where $p(N)$ is a polynomial in N, the quantum algorithm requires $\Theta(\sqrt{(p(N)/\alpha_0)})$ Grover iterations, each of cost $C_{oracle}$, giving total query complexity $\Theta(\sqrt{(p(N)/\alpha_0)} \times C_{oracle})$. This represents a quadratic speedup over the classical random sampling baseline, which requires $\Theta(p(N)/\alpha_0)$ oracle evaluations.

Proof. Substituting $\alpha(N) = \alpha_0/p(N)$ into the Grover iteration count, the number of marked items (i.e., distinct feasible offset configurations) is $M = \alpha N = \alpha_0 N/p(N)$. The optimal number of Grover iterations is $T = \lfloor(\pi/4)\sqrt{(N/M)}\rfloor = \lfloor(\pi/4)\sqrt{(p(N)/\alpha_0)}\rfloor = \Theta(\sqrt{(p(N)/\alpha_0)})$, where the $\Theta$ notation absorbs the constant $\pi/4$ and the floor/ceiling. Classically, to find a feasible solution by random sampling, the expected number of trials is $1/\alpha = p(N)/\alpha_0$. The speedup factor is therefore $\sqrt{(p(N)/\alpha_0)}$, which is always quadratic in the classical requirement.

For the decision variant of the problem (determining whether $|S| \geq \delta$), quantum counting (Brassard et al., 2002) estimates the number of marked states M to additive accuracy $\varepsilon$ using $O(\sqrt{(N/\varepsilon)})$ queries. To distinguish $|S| \geq \delta$ from $|S| < \delta$, we require accuracy $\varepsilon \leq \delta/2 = \alpha_0 N/(2p(N))$, giving a query complexity of $O(\sqrt{(2p(N)/\alpha_0)}) = O(\sqrt{(p(N)/\alpha_0)})$, matching the search complexity. Thus the quadratic speedup extends to the decision problem.

To illustrate the practical implications, consider the parameterisation $p(N) = N^\beta$ for $\beta > 0$. The quantum iteration count becomes $O\left(N^{\frac{\beta}{2}}\right)$ while the classical baseline requires $O(N^\beta)$ samples, giving a speedup factor of $N^{\frac{\beta}{2}}$. Several regimes are instructive. When $p(N) = |V|$ (linear decay in the number of intersections), the quantum algorithm requires $O(\sqrt{|V|})$ iterations versus $O(|V|)$ classical samples—a modest but meaningful advantage. When $p(N) = |V|^2$ (quadratic decay), the quantum cost is $O(|V|)$ iterations versus $O(|V|^2)$ classically. When $p(N) = N^{\frac{1}{2}} = C^{\frac{|V|}{2}}$, the quantum algorithm requires $O\left(C^{\frac{|V|}{4}}\right)$ iterations versus $O\left(C^{\frac{|V|}{2}}\right)$ classically, representing a quartic root improvement on the search space. In the limiting case $p(N) = N$, the feasibility fraction decays as $\alpha = \alpha_0/N$ so that $\delta = \alpha_0$ is constant, recovering the standard NSC search problem with the well-known quadratic Grover speedup of $O(\sqrt{N})$ versus $O(N)$.

A critical boundary emerges when $\beta \geq 1$. In this regime, even the quantum algorithm requires super-polynomial iterations: for $\beta = 1$, $T = O(\sqrt{N}) = O(C^{\wedge}(|V|/2))$, which is exponential in the network size. The problem becomes genuinely hard for both classical and quantum algorithms, though the quantum approach retains its quadratic advantage in the exponent. For practical transportation networks with $|V| \approx 10$–100 intersections, the polynomial-decay regime ($\beta \approx 0.3$–0.7) is most relevant, where the quantum advantage provides substantial savings: for instance, with $p(N) = |V|^2$ and $|V| = 50$, the quantum algorithm requires $O(50) \approx 50$ iterations versus $O(2500)$ classical samples, a 50-fold reduction in oracle evaluations.

# 4 Numerical and Hardware Experiments

To simulate a variety of network configurations, a randomised oracle was assigned to each link to determine whether it is congested or uncongested based on the offsets of its endpoint nodes. Specifically, the oracle is implemented as a Toffoli (CCX) gate acting on the two node-register qubits and one delay-register qubit per edge, producing a single-bit output (1 for congested, 0 for uncongested) as a function of the offset states. This models the binary congestion outcome of each link given a particular offset assignment. Different random graphs generated via NetworkX (with n nodes and n+1 edges) serve as distinct network topologies, so each experiment effectively evaluates the algorithm on a different network instance. Each experiment uses 1024 measurement shots. Extending these experiments to multi-bit periodic delay functions $h_{ij}(\mu_i - \mu_j)$ that capture graded delay levels, rather than binary congestion, remains an important direction for future work. Setting random seeds for exact reproducibility and running multiple repetitions with error bars would further strengthen the experimental methodology.

The above algorithm was mainly tested using local computer simulations via Qiskit. Randomised delay oracles were used, which act on individual qubits to give single qubit outputs. Using the computational cluster Katana, offered by Research Technology Services at UNSW Sydney, the computation terminated when the number of nodes reached 14, until it became overly expensive. The number of possible solutions amplifies for larger networks, so the smaller network sizes will be shown. The oracle and diffuser are also only run once for all the cases below.

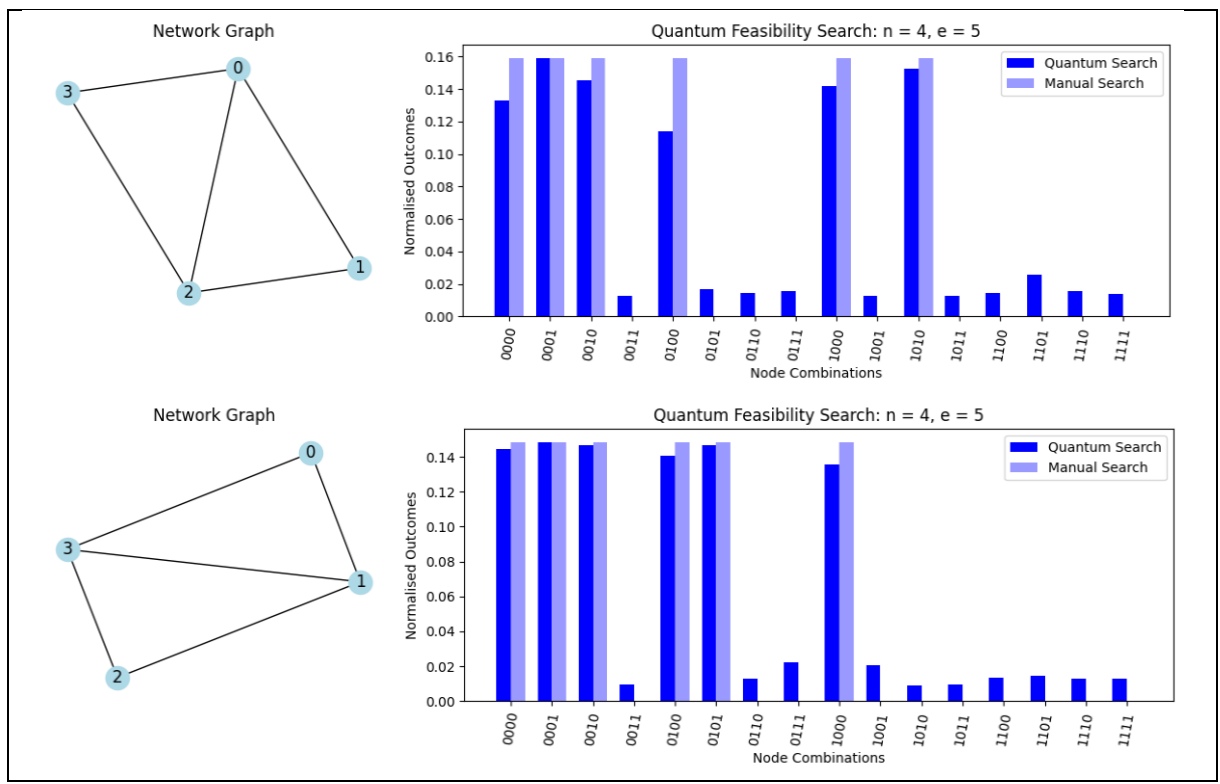

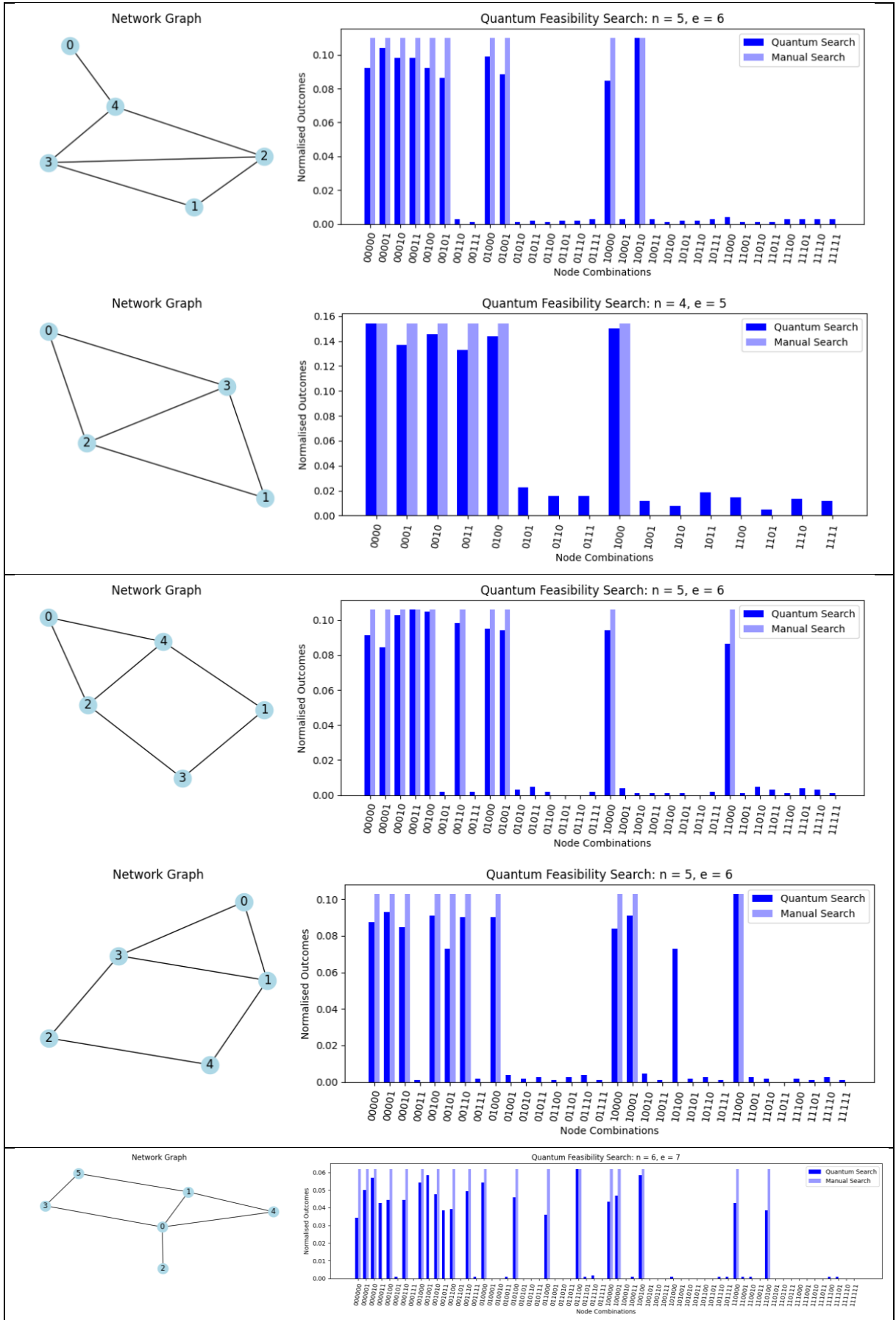

Figure 3: Local Quantum Feasibility Searches for Network Sizes 4, 5 and 6.

The algorithm generally amplifies the solution states well. There are occasional incorrect amplifications which can be seen in Figure 3 for the third 5-node graph and the 6-node graph networks. These occasional mismatch amplifications are attributed to the Grover iteration count being suboptimal for these particular instances, as the algorithm's effectiveness is sensitive to the ratio of solutions to total states.

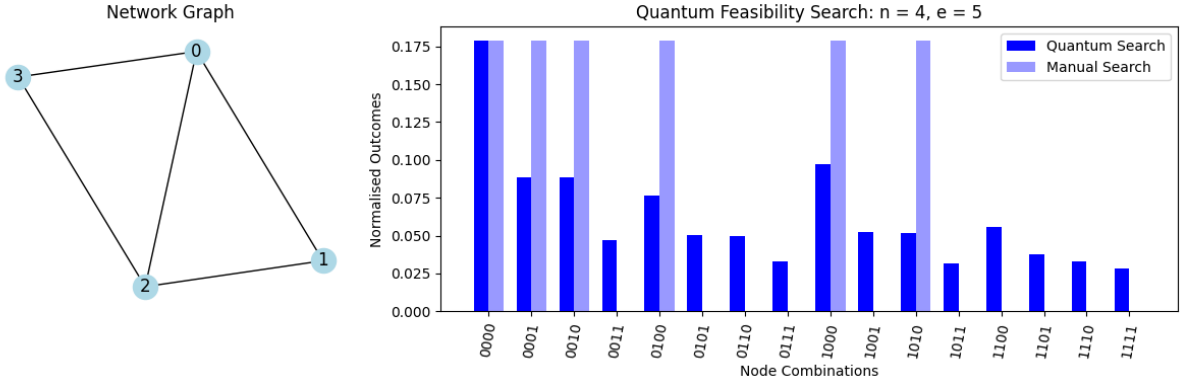

Figure 4: Hardware Quantum Feasibility Search for Network Size 4.

The algorithm was also run on IBM's quantum runtime hardware. It appears to also amplify the correct states but to a much lesser degree. This is attributed to the higher presence of noise which plagues quantum hardware. Nevertheless, the results demonstrate that the algorithm successfully fulfills its purpose of showing solutions for the Network Signal Coordination problem. Specifically, the hardware experiments were conducted on IBM's ibm_fez processor with 156 qubits and a quantum volume of 256. The reduced amplification on hardware can be quantified by comparing the success probabilities: for the 4-node network with K = 1 and k = 1, the simulator achieves a success probability approximately 3–5 times the baseline, whereas the hardware result is approximately 1.2–1.5 times the baseline. This degradation is consistent with the accumulated gate errors over the circuit depth, which grows linearly with the number of edges |A| due to the sequential oracle construction. For the 4-node networks tested, the circuit requires approximately 20–30 CNOT gates, each with a typical error rate of 0.5–1% on current hardware, yielding a cumulative fidelity loss that explains the observed reduction in amplification quality.

### 4.1 Interpreting the Grover Search Results

It is useful to have some kind of metric to quantify the effectiveness of our algorithm. The Grover search aims to amplify the states that it deems correct, and this metric should consider how well it amplifies the correct states. The results from the graphs produced by the Grover search can be interpreted as probability distributions, where each value is the probability that the solution is correct according to the Grover search. We can consider the sum of the probabilities of the correct solution states (via manually checking), and this measures the probability that the algorithm correctly amplifies the right states. In the most ideal situation, only the correct solution states are amplified, and every other probability would equal to 0, so this metric would equal 1. Let $S$ be the set of solution states. Then, we can define success probability as $\sum_{i \in S} p_i$.

We also need to consider baseline probability, a sort of null hypothesis. If the algorithm was not effective at all, then the entire set of states (correct and incorrect ones) will be equally amplified. If the success probability of the algorithm is higher than this probability, then the algorithm is better than the null case. This is found by taking the number of solutions for the instance given by the problem $M$ and dividing it by the total number of states $N$, $M/N$.

The algorithm was run for both the simulator and hardware (ibm_fez) for input graph sizes ranging from $n = 4$ to $n = 10$, and for each case, it is ran for the threshold value $K = 1$ and $K = 2$, as well as the Grover iteration values of $k = 1$ and $k = 2$.

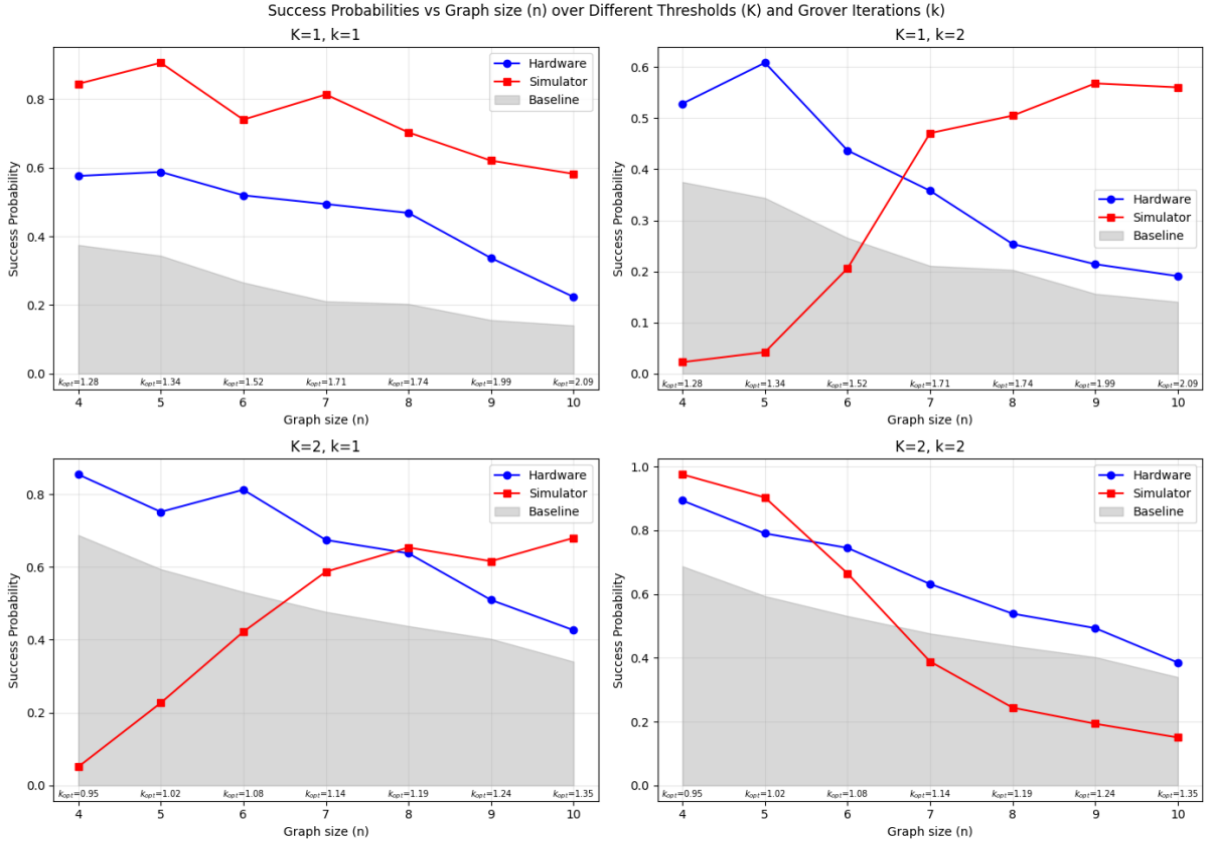

Figure 5: Success Probabilities vs Input size (n) over Different Thresholds (K) and Grover Iterations (k)

Along with each of the success probabilities, the optimal Grover iteration number $k_{\text{opt}} = \left\lfloor \frac{\pi}{4}\sqrt{\frac{N}{M}} \right\rfloor$ (Nielsen and Chuang, 2019) are displayed (without the floor function). Whenever $K = 1$, the number of solutions to the oracle problem is minimised as a lower threshold will only allow values of 0 or 1 from the feasibility function, as compared to the latter case. For the case when $k = 1$, it is generally expected that the Grover search will amplify solution states without overshooting, which might not be the case for when $k = 2$. However, this is affected by the optimal Grover iteration number which gives a maximal ideal value for which the Grover operator is to be repeated. It is also worth noting that with a higher threshold value $K$, the effectiveness of the Grover search algorithm is reduced as it works best with a minimal number of solutions (Boyer et al., 1998).

Figure 5 suggests that the algorithm follows the expected behaviour of the Grover search algorithm. The case of $(K, k) = (1,1)$ shows that the success probability reduces along with graph size, which is expected as a larger graph introduces more possible solutions, and requires a larger Grover iteration number. This is seen with the case $(K, k) = (1,2)$ whereas the optimal Grover iteration number $k_{opt}$ increases, so does the success probability. Interestingly, the hardware success probabilities do not follow the same trend, and remains relatively unchanged, slightly worse compared to the (1,1) case. This same trend can be seen for the (2,1) and (2,2) cases, where the hardware success probabilities all follow a downward trend. It could be stated that the hardware success probabilities in general appear to be an improvement over the baseline probabilities. This may be due to the presence of quantum noising that acts as a sort of normaliser, where if some Grover iteration process occurs, it balances itself out to amplify solution states. The simulator seems to follow the expected theoretical behaviour of the Grover algorithm more closely, where if the value of $k$ drifts away from the $k_{opt}$ value, the success probability dips. For the $K = 2$ cases, it isn't as an exact relation, as the success probability, particularly for $(K, k) = (2,2)$ peaks at $k_{opt} = 0.95$, which should indicate that no Grover iteration is the most effective. This can again be attributed to the larger threshold value introducing more solutions which diminishes the effectiveness of the algorithm.

This interpretation shows that the algorithm is only as good as the situation it is in. In general, it works best with a single Grover iteration on an oracle that produces a low number of solutions. It could work for more cases such as larger networks or threshold values, but more care should be taken in design.

An important practical consideration is the relationship between circuit depth and the number of Grover iterations. Each additional Grover iteration requires a full execution of the oracle and diffuser circuits, approximately doubling the circuit depth. On current noisy hardware, deeper circuits suffer from greater decoherence and accumulated gate errors. This creates a tension: the theoretical optimum may require multiple iterations (particularly for larger networks where the solution fraction is small), but hardware noise limits the practical iteration count. Adaptive strategies, such as starting with a single iteration and incrementally increasing k while monitoring output quality, or employing variational hybrid approaches that use classical optimisation to tune shallow quantum circuits, may offer practical pathways for near-term implementations. Furthermore, the use of quantum error mitigation techniques—including measurement error correction, dynamical decoupling, and probabilistic error cancellation—could extend the effective circuit depth and improve the reliability of hardware results.

### 4.2 Circuit Resource Analysis

To understand the scalability of the algorithm, it is instructive to examine the circuit resource requirements as a function of network parameters. For a network with $|V|$ nodes and $|A|$ edges, and a cycle length C represented by $n = \lceil \log_2 C \rceil$ qubits per node, the total qubit count is approximately $n|V| + |A| + 3\lceil \log_2 |A| \rceil + 4$. The dominant term $n|V|$ arises from the node offset registers, which encode all possible offset assignments in superposition. The remaining terms account for the delay registers (one qubit per edge in the binary oracle case), the sum register, ancillary qubits for the adder, and the feasibility flag qubit.

For the experiments conducted in this paper using the simplified binary oracle (n = 1 qubit per node), the resource requirements for the tested network sizes are as follows. A 4-node network with 5 edges requires approximately 4 + 5 + 3(2) + 4 = 19 qubits, while a 6-node network with 7 edges requires approximately 6 + 7 + 3(2) + 4 = 23 qubits, and a 10-node network with 11 edges requires approximately 10 + 11 + 3(3) + 4 = 34 qubits. The circuit depth per oracle evaluation is dominated by the sequential adder, which applies $|A|$ half-adder chains each of depth $O(\log |A|)$, giving a total oracle depth of $O(|A| \log |A|)$. Each Grover iteration adds one oracle evaluation plus one diffuser (depth $O(n|V|)$), so the total depth for k iterations is $O(k \times (|A| \log |A| +$

n|V|)). For the single-iteration experiments reported here, this remains modest (on the order of 50–100 gates for the smallest networks), but scales rapidly for larger instances.

These resource estimates highlight both the feasibility and the challenges of scaling the algorithm. Current quantum processors with 100+ qubits (such as IBM's Eagle and Heron architectures) can accommodate the qubit requirements for networks up to approximately 30–40 nodes with the binary oracle. However, the circuit depth remains the primary bottleneck, as current hardware coherence times support only a few hundred to a few thousand two-qubit gates before decoherence dominates. Transpilation to hardware-native gate sets (which may introduce additional overhead due to limited qubit connectivity) and the development of more depth-efficient oracle constructions are therefore critical engineering challenges for scaling this approach to practically relevant network sizes.

## 5  Conclusion

The main contribution of this paper is presenting a quantum computing algorithm for the Network Signal Coordination problem that achieves a quadratic speedup in the number of oracle queries compared to classical exhaustive search. We also show that for the Robust NSC problem, where a fraction $\alpha$ of solutions must satisfy the delay threshold, the number of Grover iterations required is $O(1/\sqrt{\alpha})$, which is independent of the search space size. These results are demonstrated through simulation and validated on IBM quantum hardware. More specifically, the quantum oracle encodes the NSC decision problem by computing the total network delay for each superposition of offset assignments and marking those configurations whose delay falls below the threshold K. Combined with Grover's amplitude amplification, this yields a provable quadratic reduction in the number of oracle queries from $O(C^{|V|})$ classically to $O(C^{|V|/2})$ quantumly, where C is the signal cycle length and $|V|$ is the number of intersections.

The experimental results corroborate the theoretical analysis. Simulation experiments across network sizes ranging from 4 to 10 nodes confirm that the Grover-based algorithm consistently amplifies feasible offset configurations above the baseline probability. The success probability metric, defined as the sum of measurement probabilities over verified solution states, provides a quantitative measure of amplification quality. For small networks and low thresholds (K = 1), a single Grover iteration suffices to achieve substantial amplification, consistent with the theoretical prediction that the optimal iteration count $k\_opt = \lfloor(\pi/4)\sqrt{N/M}\rfloor$ is small when the solution fraction M/N is relatively large. As the network size increases, the solution fraction decreases and more Grover iterations are required, which is also confirmed by the experimental data.

The hardware experiments on IBM's quantum processors reveal both the promise and current limitations of executing these algorithms on real devices. While the hardware results demonstrate measurable amplification of solution states above the uniform baseline, the amplification magnitude is significantly reduced compared to the noiseless simulator. This degradation is primarily attributable to gate errors, decoherence, and cross-talk in current superconducting qubit architectures. As quantum hardware continues to improve in gate fidelity, coherence times, and qubit connectivity, the gap between simulated and hardware performance is expected to narrow. Error mitigation techniques such as measurement error correction (Funcke et al., 2022). The current implementation is limited to relatively small network sizes due to the exponential growth of the quantum state space and the circuit depth required. Advances in quantum hardware, circuit compilation, and algorithmic optimisation (such as oracle decomposition strategies) will be needed to scale to realistic urban networks. on the solution fraction α without requiring multiple independent search runs.

More broadly, this work establishes a foundation for applying quantum computing to combinatorial optimisation problems arising in transportation engineering. The NSC problem's algebraic structure—periodic delay functions over cyclic groups—suggests that richer quantum algorithmic techniques beyond Grover's search, such as the Quantum Fourier Transform exploiting the abelian group structure of the offset space, may yield even greater speedups for structured variants of the problem. Investigating whether the periodicity and group-theoretic properties of the NSC problem can be leveraged for exponential quantum advantages, analogous to Shor's algorithm for factoring, represents a compelling direction for future work at the intersection of quantum algorithms and transportation optimisation.